\documentstyle[preprint,aps]{revtex}
\begin{document}
\title{Minimal SU(5) Resuscitated by Higgs Coupling Fixed Points} 
\author{P. Q. Hung}
\address{Dept. of Physics, University of Virginia, Charlottesville,
Virginia 22901}
\date{\today}
\maketitle
\begin{abstract}
The issue of gauge unification in the (non-supersymmetric) Standard
Model is reinvestigated. It is found that with just an additional
fourth generation of quarks and leptons, $SU(3) \otimes SU(2) \otimes
U(1)$ gauge couplings converge to a common point $\sim 2.65 \times
10^{15}$ GeV ($\tau_p \sim 10^{34 \pm 1}$ years) provided the Higgs boson
has a mass of {\em at least} 210 GeV. 
The presence of ultraviolet
fixed points for the Yukawa and Higgs quartic couplings is found to
be the origin of such unification.
\end{abstract}
\pacs{11.10.Hi, 12.10.Dm, 12.10.Kt}

\narrowtext

It is without any doubt that the search for the Higgs boson is one
of the most important endeavor in particle physics. Its discovery would
certainly reveal a great deal about the mechanism of symmetry breaking
and would probably provide a window into new physics beyond the Standard
Model (SM).

It is amazing that sometimes a mere knowledge of the Higgs mass can go
a long way in giving us a glimpse of such new physics if it exists. 
For instance, if the Higgs boson were
to be found at LEPII (i.e. $m_H \leq m_Z$), vacuum stability arguments suggest
that there must be new physics of some sorts in the TeV region
\cite{stability}. On the
other hand, if the Higgs boson were heavier than $2 m_Z$, one would see
the appearance of Landau poles (or possibly large ultraviolet fixed
points)  in the Higgs quartic coupling
below the Planck scale \cite{hung2}. In addition, this range ($>2 m_Z$) would 
almost certainly rule
out a general class of Supersymmetric SM (SSM) since these models predict
the existence of a light Higgs scalar of no more than 150 GeV in mass
\cite{pokorski}. (The
upper bound is even more stringent for the Minimal SSM (MSSM) and is
approximately 130 GeV \cite{hung2}.) For $m_H$ between 150 and 180 GeV,
the SM is perfectly respectable all the way up to the Planck mass despite
the fact that there remains many unanswered fundamental questions. (This
is probably the only window where ``nothing happens'', if one may say so.)
In this note, we would like to explore further the issue of the Landau poles
and ultraviolet fixed points, and,
in particular, their effects on the {\em evolution} of the gauge couplings.

Despite the absence of a direct experimental evidence for proton decay,
the idea of a Grand Unification of all known interactions is 
still very attractive for a number of reasons such as the ones
espoused in the classic SU(5) paper \cite{GUT} for example.
In the simplest of such schemes, one usually assumes a ``desert'' beyond the
electroweak scale with the unification scale being determined by the
point where the three SM couplings meet.
However, it is a standard lore that present measurements of the gauge couplings
at the Z mass appear to indicate that
all three couplings actually do not meet at the
same point and this is somewhat problematic for the
simple idea of Grand Unification, such as minimal SU(5) \cite{susy}. It is
also a standard lore that with low-energy supersymmetry broken 
at around 1 TeV or less,
such a unification is possible and occurs at a desirable energy scale $
\sim 10^{16}$ GeV \cite{susy}. 
This is, of course, a remarkable result. However, if no light scalar is
found below 150 GeV, one would have to reassess the whole
issue of gauge coupling unification within the context 
of low-energy supersymmetry. We would
like to point out in this note that, whether or not the
non-supersymmetric SM gauge couplings converge to the same point,
the question really depends on, or
rather is complicated by, the mass of the Higgs boson. In particular,
for a sufficiently heavy Higgs boson ($\geq 174$ GeV)\cite{hung2}, the mere existence
of the Landau poles might drastically affect the evolution of the gauge
couplings. Simply stated:
Can there be Grand Unification with a ``desert''  if the Higgs
boson is found to be heavier than 174 GeV? Put it in another way, a
``heavy'' Higgs ($\geq 174$ GeV) scenario might be an alternative
to MSSM as far as coupling constant unification is concerned.

The issue mentioned above is of critical importance for, if the Higgs
boson were not found below $2 m_Z$ ruling out supersymmetry (or at
least the simplest version of it), it pertains to the whole question
of whether or not there will be some form of grand unification 
at all and, if so,
whether or not it is simple (with a ``desert'') or more complex 
with many intermediate scales.

Of particular importance is the feasibility
of the search for proton decay, our {\em only} direct
evidence of the idea of Grand Unification. The current
prediction of supersymmetric GUT for the proton lifetime is 
approximately $10^{37}$ years, which
puts it way beyond any foreseeable future search. (It
may be a moot point if there is no light Higgs below 150 GeV.) The well-known prediction
of minimal SU(5) for the proton lifetime is roughly two orders of
magnitudes lower than the current experimental lower bound of $5.5 \times 10^{32}$
years \cite{pdecay}. Is it possible that, if the proton does decay, its lifetime might be
within reach of, say, SuperKamiokande which presumably could extend its search up
to $10^{34}$ years? We would like to point out in this note that this might be possible.

First let us briefly recall how ``unification'' is accomplished in MSSM. Basically, 
the extra degrees of
freedom coming from supersymmetry affect the evolution of the gauge
couplings in the following way: the slowing-down of the QCD coupling and,
because of the need for two Higgs doublets, the enhanced asymmetric
running of the $U(1)_Y$ and $SU(2)_L$ gauge couplings. The net result
is the convergence of all three couplings at an energy scale of
approximately $10^{16}$ GeV.

For the non-supersymmetric SM, we shall distinguish two cases: the minimal SM
with three generations and one Higgs doublet (case I), and the ``non-minimal''
case with four generations and one Higgs doublet (case II). We shall leave out
multiple Higgs scenarios for future studies.

In what follows, we shall use two-loop RG equations for the two cases. For
case I, they are well-known and the explicit expressions can be found in
the literature \cite{twoloop1}. For the second case with four generations, we shall
write down explicitely the two-loop RG equations below \cite{twoloop2}. 
To set the notations
straight, our definition of the quartic coupling in terms of the Higgs mass
is $m_H^2 = \lambda v^2 /3$ (corresponding to $\lambda (\phi^{\dagger} \phi)^2/6$
in the Lagrangian) while the more common definition is $m_H^2 = 2\lambda v^2$.
Therefore our $\lambda $ is {\em six} times the usual $\lambda$. The RG equations given
below reflect our convention on the quartic coupling. There is a reason
for paying close attention to this: at tree level, the value of the usual $\lambda$
where the Higgs self-interactions become strong is $8 \pi /3$.
In our case, this would be $\lambda = 16 \pi$. We shall use this value
as a guide- and only as a guide- for the evolution of the couplings,
noticing that $16 \pi$ is the value at {\em tree level}. The breakdown
of the perturbation expansion might even occur at $16 \pi^2$ instead of
$16 \pi$. In fact, with our definition of $\lambda$, the tree-level 4-point
function goes like $\lambda$ while the one-loop 4-point function 
goes like $(\lambda^{2}/32 \pi^2)(log)$. Naively one would expect the one-loop
term to be of the order of the tree-level one when $\lambda \approx
32 \pi^2$. The tree-level unitarity limit of $16 \pi$ is a factor of 6
less than this naive expectation. We will keep this in mind when the question
arises as regards to when the perturbation expansion breaks down. 

We first show that, for the minimal case with three generations and one
Higgs doublet, the numerical solutions to the two-loop renormalization
group (RG) equations reveal the following features. 1) For $m_H \geq
174 $ GeV, the energy scales where $\lambda/16 \pi \geq 1$ appear below
the Planck scale. This has been extensively studied earlier by various
authors (e.g. Ref.\cite{hung2}) and these scales are 
usually associated with the so-called ``Landau poles''.
If there were no ultraviolet fixed points, the Landau singularity is expected
to appear soon after that scale. However the two-loop RG equations are
coupled and something very different happens, which brings us to the 2nd point.
2) $\lambda$ does not actually blow up but tends towards some kind of
``ultraviolet'' fixed point, $\lambda_{fixed} \approx 79$, regardless of its original
``low energy'' value as long as $m_H \geq 180 $ GeV. Now this value of 79
for the fixed point obtained from the two-loop RG equations
is not too different from the tree-level unitarity limit of $16 \pi$.
Taking into account the possibility that one might be able to evolve $\lambda$
up to $32 \pi^2 \approx 316$ before perturbation theory breaks down, one might assume that
one can still use the two-loop RG equations to evolve the remaining couplings
beyond the scale where this fixed point is first reached.
Although the existence
of this fixed point arises from the two-loop RG equations, we conjecture that
it exists at approximately the same value to all
orders. In fact, it is well-known that the
first two terms of the $\beta$-function is 
{\em renormalization-scheme independent} and, therefore, have physical
significance.
The coefficients of the higher loop terms can be small or large
depending on a particular renormalization scheme.
3) The evolution of the top Yukawa couplings and the gauge couplings
is affected very little by the Higgs quartic coupling as it reaches
its fixed point and remains there, if the two-loop approximation 
can be trusted. As a result,
the three SM gauge couplings {\em do not seem to converge}. This
reconfirms the standard result (although previously it was not clear how the 
Higgs mass might affect this result). We found that
the energy scales where $g_2$ and $g_1$ meet and
where $g_3$ and $g_2$ meet differ by three orders of magnitude, for the range
of Higgs boson masses considered. Explicitely, for $m_H \geq 180$ GeV,
$g_2$ and $g_1$ meet at $\sim 6 \times 10^{13}$ GeV, while $g_3$ and $g_1$
meet at $\sim 4 \times 10^{14}$ GeV and $g_3$ and $g_2$ meet at $\sim 1 \times
10^{16}$ GeV.

This little exercise above was meant to show that, in the two-loop
approximation, the non-supersymmetric SM model, with three generations and 
one Higgs doublet, appears to fail to unify the three gauge couplings even if
the Higgs boson is heavy enough to exhibit ultraviolet fixed points at 
energies below the Planck
mass. Nevertheless, one should exercise extreme caution when one of
the couplings- $\lambda$ in this case- grows large. 

We now turn to the second scenario with four generations and one Higgs
doublet. These four generations fit snugly into $\bar{5} + 10$ representations
of SU(5), except for the right-handed neutrinos which we should need if
we were to give a mass to the neutrinos. This could be
incorporated into a 16-dimensional representation of SO(10) which
splits into $\bar{5} + 10 +1$ under SU(5). We could then have a pattern of
symmetry breaking like $SO(10) \rightarrow SU(5)$ for example. These details
are however beyond the scope of this paper.

The appropriate two-loop RG equations are given by:
\begin{mathletters}
\begin{eqnarray}
16 \pi^{2} \frac{d\lambda}{dt} =&& 4 \lambda^{2} + 4 \lambda( 3 g_{t}^{2}+
6 g_{q}^{2} + 2 g_{l}^{2}-2.25 g_{2}^{2}-0.45 g_{1}^{2})\nonumber \\
&&-12( 3 g_{t}^{4} + 6 g_{q}^{4} + 2 g_{l}^{4})
+(16 \pi^{2})^{-1}[\lambda (180 g_t^{6}\nonumber \\ 
&&+288 g_q^{6}+ 96 g_l^{6} -3 g_t^{4} - 6 g_q^{4}
- 2 g_l^{4} + 80 g_3^{2} (g_t^{2}\nonumber \\
&&+ 2 g_q^{2}))-\lambda^{2} (24 g_t^{2} + 48 g_q^{2} + 16 g_l^{2})-(52/6)
\lambda^{3}\nonumber \\
&&-192 g_3^{2}( g_t^{4} + 2 g_q^{4})]
\end{eqnarray}
\begin{eqnarray}
16 \pi^{2} \frac{d g_t^{2}}{dt} =&& g_t^{2} \{9 g_t^{2} +12 g_q^{2} + 4 g_l^{2}
-16 g_3^{2}-4.5 g_2^{2}-1.7 g_1^{2}\nonumber \\
&&(8 \pi^{2})^{-1}  [1.5 g_t^{4}-2.25 g_t^{2}(6 g_q^{2}+ 3 g_t^{2}
+ 2 g_l^{2})\nonumber \\
&&-12 g_q^{4}- (27/4) g_t^{4} - 3 g_l^{4}+ (1/6) \lambda^{2} +g_t^{2}\nonumber \\
&&(-2 \lambda + 36 g_3^{2})-(892/9) g_3^{4}] \} 
\end{eqnarray}
\begin{eqnarray}
16 \pi^{2} \frac{d g_q^{2}}{dt} =&& g_q^{2} \{6 g_t^{2} +12 g_q^{2} + 4 g_l^{2}
-16 g_3^{2}-4.5 g_2^{2}-1.7 g_1^{2}\nonumber \\
&&(8 \pi^{2})^{-1}  [3 g_q^{4}-g_q^{2}(6 g_q^{2}+ 3 g_t^{2}
+ 2 g_l^{2})\nonumber \\
&&-12 g_q^{4}- (27/4) g_t^{4} - 3 g_l^{4}+ (1/6) \lambda^{2} +g_q^{2}\nonumber \\
&&(-(8/3) \lambda + 40 g_3^{2})-(892/9) g_3^{4}] \} 
\end{eqnarray}
\begin{eqnarray}
16 \pi^{2} \frac{d g_l^{2}}{dt} =&& g_l^{2} \{6 g_t^{2} +12 g_q^{2} + 4 g_l^{2}
-4.5 (g_2^{2}+ g_1^{2})\nonumber \\
&&(8 \pi^{2})^{-1}  [3 g_q^{4}-g_q^{2}(6 g_q^{2}+ 3 g_t^{2}
+ 2 g_l^{2}) -12 g_q^{4}\nonumber \\
&&- (27/4) g_t^{4} - 3 g_l^{4}+ (1/6) \lambda^{2} -(8/3)
\lambda g_l^{2}] \} 
\end{eqnarray}
\begin{eqnarray}
16 \pi^{2} \frac{d g_1^{2}}{dt} =&&g_1^{4} \{ (163/15)+(16 \pi^{2})^{-1}[
(787/75) g_1^{2} + 6.6 g_2^{2}\nonumber \\
&&(352/15) g_3^{2}-3.4 g_t^{2}-4.4 g_q^{2}-3.6 g_l^{2}] \}
\end{eqnarray}
\begin{eqnarray}
16 \pi^{2} \frac{d g_2^{2}}{dt} =&&g_2^{4} \{ -(11/3)+(16 \pi^{2})^{-1}[
2.2 g_1^{2} + (133/3) g_2^{2}\nonumber \\
&&32 g_3^{2}-3 g_t^{2}-3 g_q^{2}-2 g_l^{2}] \}
\end{eqnarray}
\begin{eqnarray}
16 \pi^{2} \frac{d g_3^{2}}{dt} =&&g_3^{4} \{ -(34/3)+(16 \pi^{2})^{-1}[
(44/15) g_1^{2} + 12 g_2^{2}\nonumber \\
&&-(4/3) g_3^{2}-4 g_t^{2}-8 g_q^{2}] \}
\end{eqnarray}
\end{mathletters}
In the above equations, we have assumed for the fourth family, for simplicity, 
a Dirac neutrino mass and, in order to satisfy the constraints of electroweak 
precision measurements, that both quarks and leptons are degenerate $SU(2)_L$
doublets. The respective Yukawa couplings are denoted by $g_q$ and $g_l$. 
Also, in the evolution of $\lambda$ and the Yukawa couplings, we have neglected,
in the two loop terms, contributions involving 
$\tau$ and bottom Yukawa couplings as well as
as the electroweak gauge couplings, $g_1$ and $g_2$. For the range of Higgs
and heavy quark (including the top quark) masses considered in this paper,
these two-loop contributions are not important to the evolution of $\lambda$
and the Yukawa couplings.

In what follows, we shall assume that, whatever mechanism (a right-handed
neutrino in this case) that is responsible for giving a mass to at least the 4th
neutrino will not affect the evolution of the three SM gauge couplings.
Also there are reasons to believe that this 4th generation might be rather
special, distinct from the other three and having very little mixing with them.
The physics scenario behind the 4th neutrino mass might be quite unconventional.

We shall assume, in this paper, that the 4th generation quarks are at least as
heavy as the top quark and the leptons are heavier than $m_Z$. The assumption
concerning the 4th generation quarks is for mere convenience. (As of now, there
does not appear to be any strict limit on the masses of those quarks if the
4th family is {\em non sequential}, i.e. possibly having very little mixing
with the other three.) For the leptons, we just take LEPII as a guidance for
possible masses.

As with the case of three generations, one has two options: 1) Stop
evolving {\em all} couplings as soon as one or more couplings ($\lambda,
g_{i}^{2}$, $i= t,q,l$) reaches
$4 \pi$,or 2) Evolve the couplings beyond these values
to see if there are any ultraviolet fixed points. Option 2 is what
we will adopt in this paper for the following reason: if the fixed point
(in the two-loop approximation) is located at a not-ridiculously high value
then one might try to see what effects it has on the evolution of the
gauge couplings, just as we have done above with case I. 

We found the following results. First, for a given 4th generation quark mass ($m_Q$)
as well as 4th generation lepton mass ($m_L$), there is a minimum value of the Higgs mass
which arises because of the requirement that $\lambda >0$ (vacuum stability). As
the Higgs mass exceeds that given lower bound, an ultraviolet fixed point appears.
We have observed this behaviour for a large range of 4th generation masses.
We found the common ultraviolet fixed points to be: $\lambda \approx 108$,
$g_t^{2} \approx 28$, $g_q^{2} \approx 56$, and $g_l^{2} \approx 55$, regardless
of the initial values as long as there is an appearance of a fixed point.
It is found that these fixed points appear rather ``early'' ($\sim 1-5 
\times 10^{10}$ GeV), which means
that they have ``time'' to influence the evolution of the gauge couplings. It
is also found that the ultraviolet fixed points are quite ``stable'' all the
way to the unification scale.

The first set of 4th generation masses that we looked at is the following:
$m_Q = m_t$ and $m_L = m_Z$ (experimentally motivated choices). This set of values
is first chosen so we can find the {\em lowest} Higgs mass where our scenario would
work. We found $m_{H, min} \approx 210$ GeV. The three gauge couplings are found
to converge to a common value $\alpha_G^{-1} \approx 37.6$ at the scale
$M_G \approx 2.55 \times 10^{15}$ GeV. This would translate into a proton
lifetime (within SU(5)) of $\sim 10^{34 \pm 1}$ years, a value which is quite 
welcome from an experimental viewpoint.

The above result indicates that convergence of the three gauge couplings is
achieved, for $m_Q \geq m_t$ and $m_L \geq m_Z$, only when the Higgs boson
is heavier than $\sim$ 210 GeV. As we vary the 4th generation masses, the
lower bound on the Higgs mass will also increase. Notice that the minimum Higgs
mass is {\em always} larger than the 4th generation masses.

Since it is impossible to list all possible masses for the 4th generation, we
shall illustrate our scenario with another two sets of values. The results
are listed in Table 1. In practice, one can deduce the lower bound on the
Higgs mass once the 4th generation quark and lepton masses are known. If
for some reason, the Higgs boson and one set of fermions (quarks or leptons)
are known first, one can set an {\em upper limit} on the remaining set of fermions.
The RG equations listed above can be easily used for such purposes. It is
beyond the scope of this paper to carry out such a task.

We obtain the same results for widely different
Higgs masses, for a given set of 4th generation masses. This comes about because
the same fixed point is reached for any $\lambda$ above its minimal value. There
is an attractive domain lying above the minimal critical value. For convenience
only the minimum masses for the Higgs boson are listed in Table 1.

As can be seen from Fig. 1, the three gauge
couplings do actually converge at a scale of approximately $2.65 \times
10^{15}$ GeV and at a value $\alpha_G \approx 1/39$. (These numbers 
correspond to values of 4th generation masses as shown in
Table 1.) Not only do they meet
but also at a comfortable mass scale. Note that $\alpha_G$ is practically
the same as the ``old'' value with the only difference being the 
``unification'' scale. It is found that $\alpha_G$ is very weakly dependent
on the initial Higgs mass, varying between 1/38 and 1/39 for a large
range of masses. It is beyond the scope of this paper to illustrate this
minor variation.

The proton partial mean lifetime as represented by
$\tau_{p \rightarrow e^{+} \pi^{0}}$ is predicted to be
$\tau_{p \rightarrow e^{+} \pi^{0}}(yr) \approx 10^{31\pm 1}(M_{G}/4.6 \times
10^{14})^{4}$. In our case, we obtain the following prediction:
$\tau_{p \rightarrow e^{+} \pi^{0}}(yr) \approx 1 \times 10^{34\pm 1}$.
This is comfortably larger than the current lower limit of $5.5 \times 10^{32}$
years. In addition, the prediction is not too much larger than the
current limit which means that it might be experimentally accessible in
the not-too-distant future, in contrast with the MSSM predictions. 

This scenario made a number of predictions: 1)the proton decays at an accessible
rate $\sim 10^{34 \pm 1}$ years; 2) there is a fourth generation of quarks and leptons;
3) For given 4th generation quark and lepton masses, there is a lower bound on the
Higgs boson mass above which the quartic and Yukawa couplings will evolve into
an ultraviolet fixed point (with the consequences discussed above). All of these
features can be tested in a not-too-distant future. For example, the fourth generation
can be {\em non-sequential} and can have exceptionally long lifetimes. This
could provide a distinct signature\cite{frampton}. Last but not least, the Higgs
boson, if found, is predicted to be at least 210 GeV and possibly heavier,
depending on the 4th generation masses, for unification to occur.

I would like to thank Paul Frampton and Gino Isidori for helpful discussions and
comments on the manuscript.
I would also like to thank the theory groups at the University of Rome ``La Sapienza''
and at the Ecole Polytechnique, Palaiseau, for the warm hospitality where part of
this work was carried out. This work is supported in parts by the US Department
of Energy under grant No. DE-A505-89ER40518.

%
%
\begin{figure}
\caption{The evolution of the SM gauge couplings squared versus ln(E/175 GeV). $g_3$,
$g_2$, and $g_1$ are the couplings of SU(3), SU(2) and U(1) respectively. 
We have used $g_3^{2}(m_t) = 1.396$, $g_2^{2}(m_t) =0.418$, and
$g_3^{2}(m_t) = 0.2115$. The point
where the couplings meet corresponds to a unification scale of $\sim 2.65 \times
10^{15}$ GeV. The common unified coupling is $\alpha_G \sim 1/39$.}
\end{figure}

%
%
\begin{table}
\caption{The minimum value of the Higgs mass for two sets of 4th generation
masses. Here $m_t = 175$ GeV. The corresponding couplings run to a fixed point:
$\lambda \approx 108$, $g_t^{2} \approx 28$, $g_q^{2} \approx 56$, and 
$g_l^{2} \approx 55$. Larger values of Higgs masses also lead to the same fixed
point, with the same consequences.}
\begin{tabular}{lcc}
$m_{H,min}$(GeV) & 224 & 233\\
$m_{Q}$(GeV) & 180 & 180 \\
$m_{L}$(GeV) & 135 & 156
\end{tabular}
\end{table}

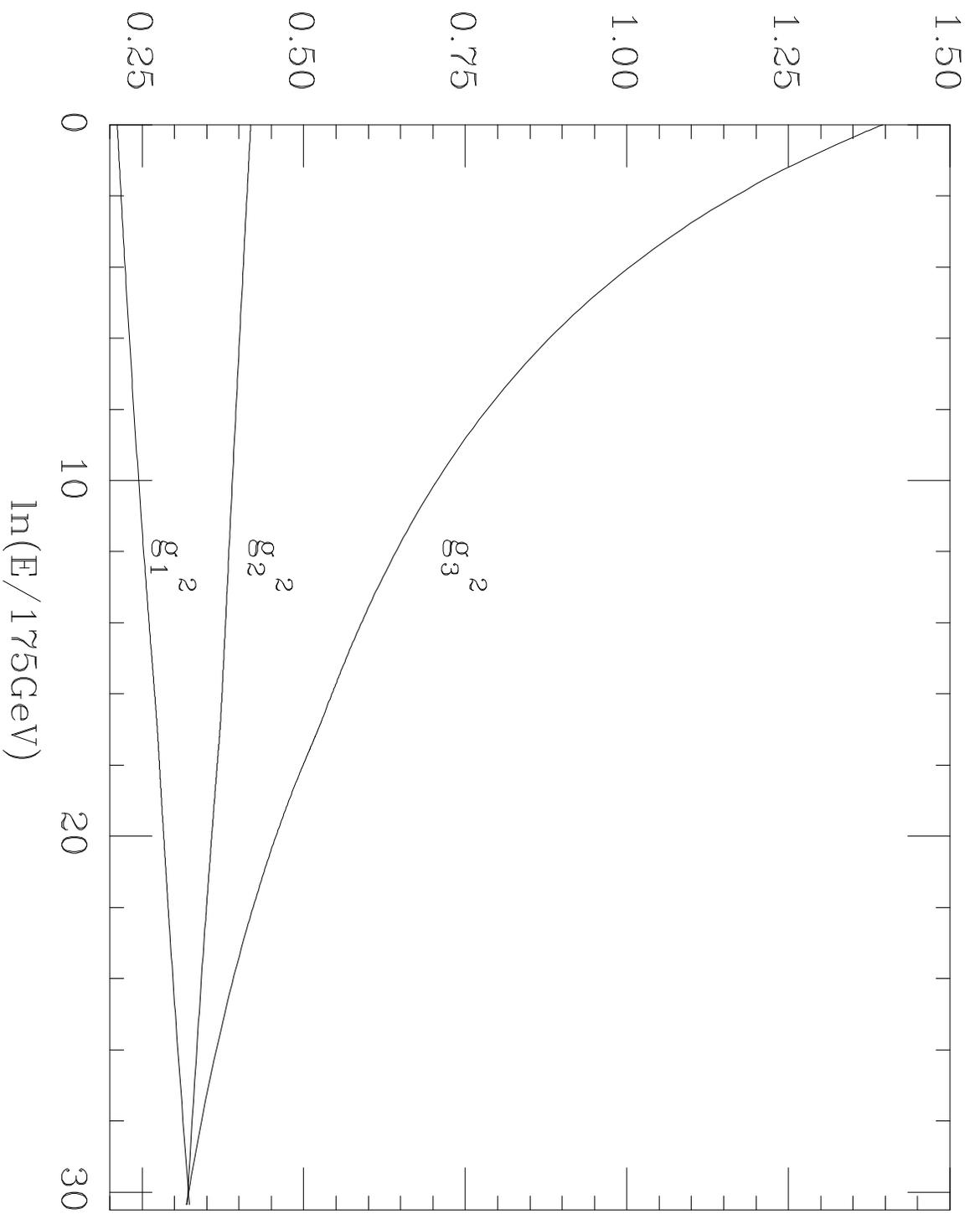
\end{document}